\documentclass[aps,prl,twocolumn,groupedaddress, showpacs]{revtex4-1}
\usepackage{graphicx}
\graphicspath{{./figs/}}
\usepackage{notes2bib}
\bibnotesetup{
note-name = ,
use-sort-key = false
}

\begin{document}

\title{Picosecond Dynamic Heterogeneity, Hopping and Johari-Goldstein Relaxation in Glassforming Liquids}

\author{Marcus T. Cicerone}
\email[]{cicerone@nist.gov}
\author{Qin Zhong}
\author{Madhusudan Tyagi}

\affiliation{National Institute of Standards and Technology, Gaithersburg, MD 20899-8543}

\date{\today}

\begin{abstract}
We show that incoherent quasi-elastic neutron scattering (QENS) from molecular liquids reveals a two-state dynamic heterogeneity on a 1 ps timescale, where molecules are either highly confined or are free to undergo relatively large excursions. Data ranging from deep in the glassy state to well above the melting point allows us to observe temperature-dependent population levels and exchange between these two states. A simple physical picture emerges from this data, combined with published work, that provides a mechanism for ``hopping" and for the Johari-Goldstein ($\beta_{JG}$) relaxation, and allows us to accurately calculate the diffusion coefficient, $D_T$, and characteristic times for $\alpha$, and $\beta_{JG}$ relaxations from ps timescale neutron data.
\end{abstract}


\maketitle


The dynamics of liquids is not completely understood, but seems to be quite complex. In addition to viscosity and the closely related $\alpha$ relaxation, there are at least two other apparently universal relaxation processes, denoted as ``fast $\beta$" ($\beta_{fast}$) and ``Johari-Goldstein $\beta$" ($\beta_{JG}$) \cite{Johari:TheJournalOfChemicalPhysics:1970}. All three have been topics of investigation for decades, yet many questions about the individual processes and their inter-relationship are unresolved. 

It seems that each of these processes may be influenced by short-timescale dynamic heterogeneity. The $\beta_{fast}$ process, which occurs at $\approx$1 ps, has been historically considered to arise only from uniform vibrational motion, but was recently shown to also contain a signature of collective motion \cite{Russina2000}. An influence of short-time dynamic heterogeneity on $\alpha$ relaxation was suggested by Goldstein \cite{Goldstein:TheJournalOfChemicalPhysics:1969}, who proposed a thermally-induced hopping over saddle points on a potential energy landscape (PEL). Mode coupling theory (MCT) also seems to require a phenomenon such as hopping to properly account for $\alpha$ relaxation below a critical temperature ($T_c$)\cite{Gotze:ZeitschriftFurPhysikBCondensedMatter:1987}. The connection of the $\beta_{JG}$ process to short-time dynamic heterogeneity is less well established, but evidence for a connection seems to be building \cite{Johari:JournalOfNonCrystallineSolids:2002, Goldstein:JCPComm:2010}. In particular, $\beta_{JG}$ relaxation appears to involve interbasin transitions in the PEL formalism, similar to $\alpha$ relaxation. It is argued that the former should thus also feel influence from dynamic heterogeneity \cite{Goldstein:JCPComm:2010}. 

We present analysis of incoherent quasielastic neutron scattering (QENS) that provides a molecular mechanism, rooted in ps timescale dynamic heterogeneity, for the $\beta_{JG}$ process and its relation to hopping and to $\alpha$ and $\beta_{fast}$ relaxation. 

\begin{figure}[htbp]
\begin{center}
\includegraphics[width=9cm]{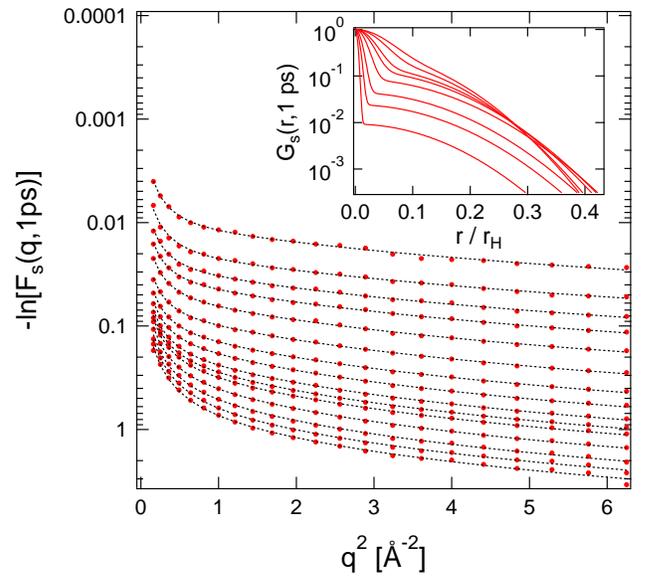}
\caption{ISF at 1 ps for propylene glycol (PG) at T = 60, 90, 120, 150, 180, 210, 240, 265, 290, 307, 320, 350, 375, 400, and 425 K (circles, top to bottom), with fits to Eq (1) (dashed lines). Inset: Van Hove function at 1 ps corresponding to fits at T = 60, 120, 180, 240, 290, 320, 375, and 425 K (bottom to top). Data were collected at the NIST neutron center on NG4 \cite{Copley:ChemicalPhysics:2003} with an energy resolution of $200 \mu$ eV \& $\lambda = 4.0$ \AA.\label{VanHove}}
\end{center}
\end{figure}

We have performed QENS on five liquids, propylene carbonate (PC), propylene glycol (PG), glycerol, ortho-terphenyl (OTP) and sorbitol. $S(q,\omega)$ data from each material was obtained over a momentum transfer range of (0.2 to 2.5) \AA$^{-1}$ and energy transfer range of (0.19 to 4.5) meV, and was transformed to F(q,t) for fitting. Care was taken to avoid potential artifacts due to multiple scattering and crystallization; this, along with fitting details are discussed in the Supplementary Material \bibnote[SI]{See Supplemental Material at [URL] for details on sample and data handling.}.  Figure \ref{VanHove} shows intermediate scattering function (ISF,F(q,t)) data from propylene glycol at 1 ps (approximate timescale for $\beta_{fast}$ relaxation), and fits to a two-state model \cite{Chudley:ProceedingsOfThePhysicalSociety:1961,Teixeira:PhysRevA:1930}: 
\begin{equation}
F_{s}(q) = (1-\Phi) e^{(-\pi^{2} \sigma_{TC}^{2} q^{2})} + \Phi e^{(-\pi^{2} \sigma_{LC}^{2} q^{2})} \label{2Gauss}
\end{equation} 
where $\sigma_{TC}$ and $\sigma_{LC}$ indicate characteristic lengthscales of motion, and $\Phi$ represents the fraction of molecules exhibiting motion characterized by $\sigma_{LC}$. The parameters $\sigma_{LC}$ and $\sigma_{TC}$ have very low co-variance \bibnote[SI]. because their fit values are well separated.  Further, $\sigma_{LC} < q_{m} < \sigma_{TC}$, where $q_{m}$ is the peak in the structure factor, $\approx 1.4\,$ \AA$^{-1}$ for these liquids. Thus, $\sigma_{LC}$ and $\sigma_{TC}$ respectively describe intramolecular and highly localized motion. The two distinct lengthsales of motion are easily visualized in the single particle van Hove correlation function calculated from the fit parameters and shown in the inset to Fig. 1 as $G_{s}(r) = (1-\Phi) e^{(-r^2/ \sigma_{TC}^{2})} + \Phi e^{(-r^2/ \sigma_{LC}^{2})}$. Thus, we observe that the $\beta_{fast}$ process has two components. One is a localized, vibration-like motion of tightly caged (TC) molecules, and the other is a relaxation executed by more loosely caged (LC) molecules. The latter have been associated with collective motion through the q-dependence of the coherent structure factor, $S(q,\omega)$ \cite{Russina2000}.

\begin{figure}[htbp]
\begin{center}
 \includegraphics[width=9cm]{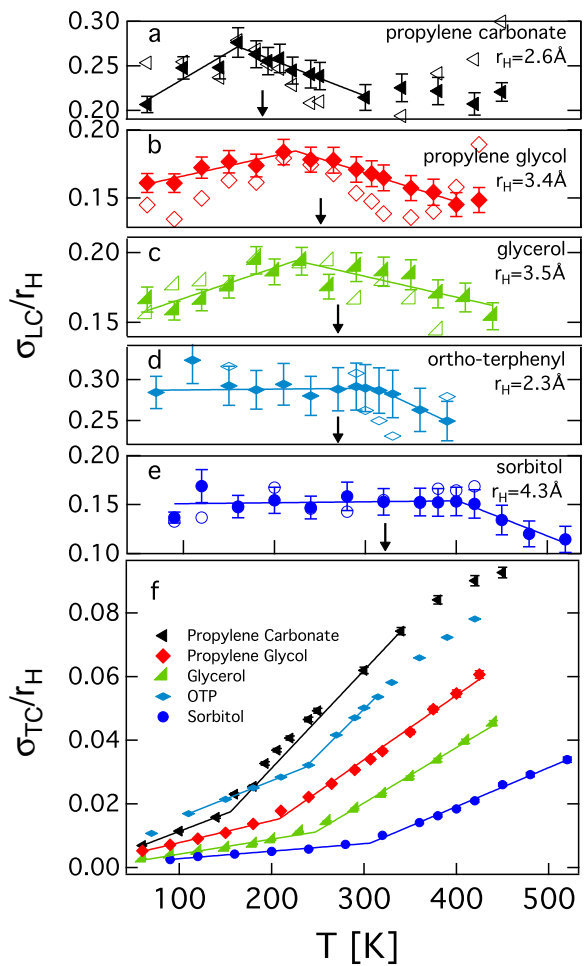}
\caption{Confinement lengthscales for LC (a - e) and TC (f) molecules in the two-state model at 1 ps (solid symbols) and 10 ps (hollow symbols). Solid lines are best fits. Arrows indicate $T_c$ for each material. Error bars indicate uncertainties in parameters at one standard deviation.\label{sigs}}
\end{center}
\end{figure}

Figure \ref{sigs} shows, in solid symbols, $\sigma_{TC}$ and $\sigma_{LC}$ values obtained at 1 ps and normalized by the high-temperature hydrodynamic radius ($r_H$) of each of the molecular species \cite{Qi:JournalOfChemicalPhysics:2000,Kruk:JPhysChemB:2011,Tolle:ReportsOnProgressInPhysics:2001,Yu:CGD:2003} as indicated in panels a-e. The hollow symbols are normalized $\sigma_{LC}$ values obtained at 10 ps. The magnitudes of $\sigma_{LC}$ fall in the range (0.1 to 0.3) $r_H$, consistent with fast collective motions seen in colloids \cite{Weeks:Science:2000}, simulation  \cite{Chaudhuri:PhysicalReviewLetters:2007}, and ionic systems \cite{Russina2000}. We note that the collective motions characteristic of the LC regions must occur on a timescale $\approx 1\,ps$ or less, since $\sigma_{LC,1ps}=\sigma_{LC,10ps}$ in all cases (except PC and PG at high temperature, which are likely influenced by $\alpha$ relaxation at 10 ps).

The temperature dependence of $\sigma_{TC}$ values is consistent with expectations for a localized component, changing markedly in the vicinity of $T_c$ and $T_g$. We ascribe the anomalously high $\sigma_{TC}$ values for OTP to ring libration, and note that the characteristic lengthscale of this motion is small compared to $\sigma_{LC}$, again attesting to the intermolecular nature of the latter. Unfortunately, this excess scattering prevents us from using the OTP $\sigma_{TC}$ data in the quantitative analysis at the end of this letter.

\begin{figure}[htbp]
\begin{center}
 \includegraphics[width=9cm]{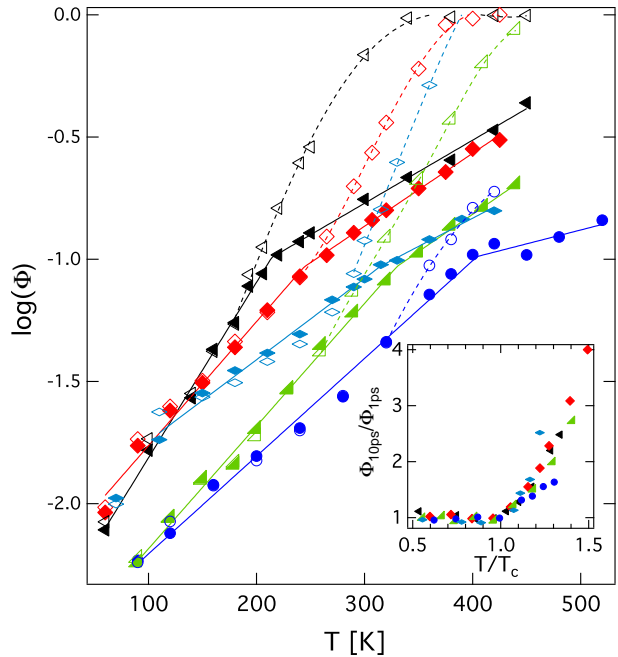}
\caption{$\Phi$ at 1 ps (solid symbols), and $\Phi$ at 10 ps (hollow symbols). Symbols have same association as in Fig. \ref{sigs}.  Solid and dashed lines are guides to the eye. The inset is described in the text.\label{Phi}}
\end{center}
\end{figure}

Figure \ref{Phi} shows $\Phi$, the fraction of molecules participating in LC states during a particular time window. The solid and hollow symbols are $\Phi$ values at (1 and 10) ps respectively for each of the liquids. At low temperatures $\Phi_{10ps}=\Phi_{1ps}$, indicating that the LC states are long-lived. On the other hand, $\Phi_{10ps} > \Phi_{1ps}$ at higher temperatures, where we can conclude that these LC domains transiently visit regions of space, allowing molecules there to undergo large excursions before the LC domain moves on, consistent with Keyes \textit{et al.} \cite{Keys:PhysicalReviewX:2011}. We observe $\Phi_{10 ps}=1$ at high temperature, indicating that all molecules are eventually involved in LC domains. The lower inset to Fig. \ref{Phi} shows that the amount of exchange between TC and LC states is significant on a the timescale of a few ps only at $T\geq T_c$.

Having established the presence of two exchangeable dynamic states, we now consider our results in the context of previous findings and formulate a physical picture of the underlying dynamics. Thermally activated \cite{Goldstein:TheJournalOfChemicalPhysics:1969} and phonon-assisted \cite{Gotze:ZeitschriftFurPhysikBCondensedMatter:1987} hopping processes were proposed, and evidence for hopping was later observed in simulation where particles were typically localized, but occasionally moved in a relatively small number of steps to distinct positions where they again became localized \cite{Miyagawa:TheJournalOfChemicalPhysics:1988}. Similar behavior is now observe routinely in simulation and model systems \cite{Weeks:Science:2000,Schroder:TheJournalOfChemicalPhysics:2000,Gebremichael:PRE:2001,Cipelletti:JournalOfPhysicsCondensedMatter:2003,Vogel:TheJournalOfChemicalPhysics:2004, Chaudhuri:PhysicalReviewLetters:2007}, and the basic mechanism for this hopping can be gleaned from results of these studies as follows: 1) The rapid excursions involve discrete cooperative rearrangements of particles from one locally preferred structure (metabasin) to another. 2.) On a ps timescale, these rearrangements generally involve only a small number of particles ($\approx$ 2 to 4), resulting in relocation by typically (0.2 to 0.3) particle radii. 3) For time $\gg$ 1 ps, cooperative motion of larger groups of particles is asynchronous, being made up of ps timescale rearrangements of smaller groups.

The correspondence between the hopping behavior seen in simulation and the behavior reported here in the LC \& TC states is clear. We observe that on a 1 ps timescale, and for $T < T_c$, most molecules are immobile, and confined to within 0.02 $r_H$. On the other hand, a small fraction (\textless 10 \%) are free to move relatively large distances, up to 0.3 $r_H$, probably through cooperative motion. Due to the time-dependent exchange between the LC and TC populations, the average molecule will be highly localized, then be transiently associated with an LC domain and freed to move away from its initial position. Subsequently it will be re-localized as the LC domain passes to a new region of space. These conditions are sufficient to yield the hopping behavior, and will \textit{necessarily} do so provided that: 1) The wait time between excursions is much longer than the time required for reorganization in an LC state ($\approx 1$ ps), and 2) Molecules make unusually large excursions as a LC domain passes through a region of space (i.e., $\sigma_{LC} > \sigma_{TC}$). Both of these conditions are manifestly fulfilled at $T < T_c$ for all the systems studied, and appear to hold even for $T < 1.5\, T_c$. 

The hopping mechanism described above requires only transient domains of rapid, cooperative motion. It is not obvious how these domains arise, but their origin must be due either to dynamics (kinetic energy fluctuations) or structural heterogeneity. A dynamic origin was initially proposed \cite{Goldstein:TheJournalOfChemicalPhysics:1969,Gotze:ZeitschriftFurPhysikBCondensedMatter:1987}. While difficult to imagine how it might physically occur\cite{Kivelson:JournalOfNonCrystallineSolids:1998}, dynamically-induced hopping was indirectly supported by a lack of correlation between structure and dynamics when the latter was averaged over a time comparable to the structural relaxation time, $\tau_{\alpha}$ at $T\geq T_c$ \cite{Berthier:PhysicalReviewE:2007,Starr:RPL:2002}. The data of Fig. \ref{Phi} suggest that dynamic states change on a timescale of a few ps at $T>T_c$, so, in retrospect it is not surprising that no correlation was found between structure and dynamics. Recent work has shown that a correlation between structure and dynamics \textit{is} found when dynamics are measured over times $< \tau_\alpha$ \cite{ottochian2011universal,Tanaka:NatureMaterials:2010}. Further, two aspects of our data suggest a predominately structural origin for hopping. Firstly, we observe a linear temperature dependence in $log(\Phi)$ (see Fig. \ref{Phi}), whereas a dynamic origin would yield a -1/T dependence from $\Phi \propto e^{-E/kT}$. Secondly, the fact that $\sigma_{LC}$ is temperature-insensitive and remains large even at 60 K strongly suggests a structural origin.

Putting aside the origin of the dynamic states, we now show that their behavior can be used to derive the key dynamic signatures of liquids. A system with two exchanging dynamic states such as discussed above may have as many as three dynamic signatures. Dynamics at the fast and slow extremes ($\beta_{fast}$ and $\alpha$) will arise from motion in the LC and TC states respectively. A third dynamic signature may arise on an intermediate timescale from exchange between the TC and LC states. Importantly, the exchange process and TC relaxation will merge if \textit{all} molecules exchange between states on a timescale comparable to or shorter than the intrinsic TC relaxation time. We propose that the intermediate timescale process, the TC-LC exchange, corresponds to $\beta_{JG}$ relaxation.

Within the proposed framework, we expect that  $\left\langle x^2\right\rangle=\sqrt{\pi} \Phi \sigma_{LC}^2/2=6D_T\tau_{\beta,JG}$ if we assume that translation occurs primarily in LC domains (since $\sigma_{LC} \gg \sigma_{TC}$), and that new displacements will occur at a rate proportional to the TC-LC exchange rate. We also expect that $\beta_{JG}$ relaxation (TC-LC exchange) will facilitate $\alpha$ relaxation when the intrinsic TC relaxation is sufficiently slow. In this limit, $\alpha$ relaxation would be facilitated by first passage of an LC domain, $log(\tau_\alpha) \propto log(\tau_{\beta,JG})/\gamma$, where $\gamma <1$ would arise from spatial correlations in the relaxation process \cite{klages2008anomalous}, due to the "stringlike" nature of the mobile particle arrangements at short time \cite{Gebremichael:PRE:2001}. We expect the TC-LC exchange to be only weakly cooperative, so treat it as a simple activated process, with activation energy, $E_a\propto1 /\sigma_{TC}$, for rearrangement of TC particles at an TC-LC interface. Under these assumptions we write expressions for these relaxation and transport processes:

\begin{equation}
\tau_{\beta, JG}=\tau_{ex}=\tau_0 \,exp\left[\frac{\delta}{k T \tilde{\sigma}_{TC}}\right]
\label{TauBeta}
\end{equation}
\begin{equation}
\frac{\tau_\alpha}{\tau_c}=\left(\frac{\Phi_c\tau_{\beta,JG}}{\Phi\tau_c}\right)^{1/\gamma}
\label{TauAlpha}
\end{equation}
\begin{equation}
D_T=\frac{\sqrt{\pi}\Phi\sigma_{LC}^2}{12\,g\,\tau_{\beta, JG}}
\label{DT}
\end{equation}
where $\tau_0$ is an inverse attempt rate, associated with $\beta_{fast}$ relaxation, $\tilde{\sigma}=\sigma/r_H$, $\tau_c$ and $\Phi_c$ are the $\alpha$ relaxation time and $\Phi$ value at $T_c$ (we assume $\tau_\alpha \approx \tau_{\beta,JG}$ at $T_c$), and g, a fitting factor, is expected to be O(1). 

\begin{figure}[htbp]
\begin{center}
\includegraphics[width=9cm]{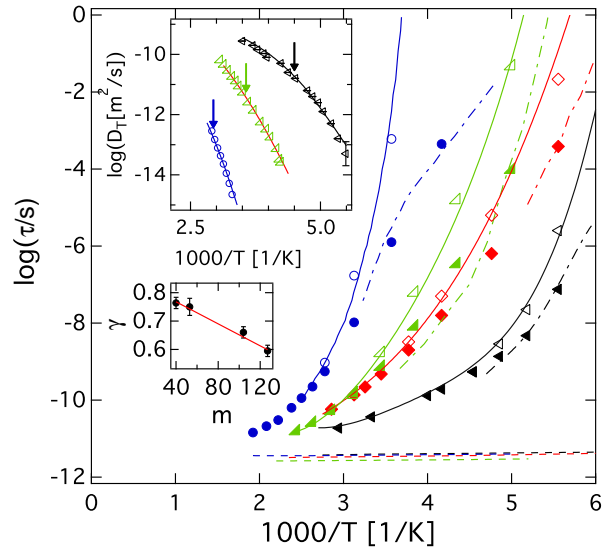}
\caption{Calculated relaxation times: $\tau_{\beta, JG}$ (solid symbols), $\tau_{\alpha}$ (hollow symbols), $\tau_0 =\tau_{\beta,fast} $ (dashed lines). Measured relaxation times: $\tau_\alpha$ (solid lines) \cite{Ngai:JCP:2001,Fujima:PhysicalReviewE:2002,borjesson:CP:1990}, $\tau_{\beta,JG}$ (dash-dot lines) \cite{Ngai:JCP:2001,Fujima:PhysicalReviewE:2002}. The upper inset shows $D_T$ values for PC \cite{Qi:JournalOfChemicalPhysics:2000}, glycerol \cite{chen:PRL:2006} and sorbitol \cite{Yu:CGD:2003} (symbols), and fits to equation \ref{DT} (solid lines). The arrows indicate $T_c$ for these liquids. Symbols have the same association with samples as in previous figures. The lower inset shows the correlation between $\gamma$ and fragility.\label{JG}}
\end{center}
\end{figure}

Figure \ref{JG} shows experimentally measured values of $\tau_\alpha$ and $\tau_{\beta,JG}$ for four of the five liquids studied here, as well as fits to Eq.s \ref{TauBeta} \& \ref{TauAlpha}. We vary $\tau_0$, $\delta$, and $\gamma$ to find simultaneously optimal fits to the $\alpha$ and $\beta_{JG}$ data. We obtain excellent fits with all measurements, except for $\beta_{JG}$  measurements of glycerol between 220 and 270 K. Those data were extracted from a weak $\beta$ peak largely buried under a strong $\alpha$ peak \cite{Ngai:JCP:2001}. Table \ref{FitPars} gives fit parameters. The parameter $\tau_0$ has values expected for $\tau_{\beta,fast}$, and we find that $\delta$ and $\gamma$ correlate strongly with the melting temperature, $T_m$, and the fragility index, $m$, respectively. We obtain the relations $\delta=-25+8.6X10^4/T_m$ with correlation coefficient $r^2=0.97$ and  $\gamma=0.85-2.0X10^{-3}m$, with $r^2=0.99$. The relationship between $\gamma$ and $m$ is shown in lower inset to Fig. \ref{JG}. 

Results from Eq. \ref{DT} are plotted in the upper inset to Fig. \ref{JG}, along with direct measurements of $D_T$ for glycerol \cite{chen:PRL:2006} and PC \cite{Qi:JournalOfChemicalPhysics:2000}, and an \textit{estimate} of $D_T$ for sorbitol \cite{Yu:CGD:2003}. Although these materials have a large variation in fragility and degree to which the Stokes-Einstein relation is violated, the model produces the correct temperature dependence for all systems for which we have diffusion data. It further gives correct absolute values within a factor of five before correction by the multiplicative fitting parameter, $g$. Additionally, the values of $g$ are very similar for the two systems for which $D_T$ has been directly measured.

\begin{table}[htdp]
\caption{Fit Parameters}
\begin{center}
\begin{tabular}{l c c c c c c c c}
& $\tau_0$ [ps] & $\delta$ [kJ/mol] &$\gamma$ &$g$ &$T_m$&m\\
\hline
PC  & 2.0$\pm$0.4 & 380$\pm$12 & 0.66$\pm$0.2&0.23$\pm$.04 &218&104\\ 
PG  & 2.5$\pm$0.4 &  377$\pm$10 &0.76$\pm$0.2& --- &214&40\\ 
glycerol & 2.5$\pm$0.4 & 248$\pm$8 & 0.74$\pm$0.3& 0.38$\pm$0.02 &291&53\\
sorbitol & 3.2$\pm$0.4 & 212$\pm$9 &0.59$\pm$0.2& 4.5$\pm$0.4 &383&127\\ 
\label{FitPars}
\end{tabular}
\end{center}
\label{fitstab}
\end{table}%

The model we propose provides an explanation for the the $\beta_{JG}$ relaxation. The model and data taken together suggest a straightforward relation between $\left\langle \tau_{\beta,JG}\right\rangle$ and $\sigma_{TC}$, which depends only on $T_m$; properly scaled with $T_m$, all the $\left\langle \tau_{\beta,JG}\right\rangle$ data will \textit{very} nearly coincide. The data further suggest that $\beta_{JG}$ and hopping arise from a structural phenomenon such as frustrated packing, indicating that these should occur in any non-crystalline liquid or solid.  

The model and data also suggest that $\alpha$ relaxation derives primarily from the $\beta_{JG}$ process at the temperatures we have investigated. At $T > T_c$, more than 10\% of molecules are involved in LC domains at any time, and, since these domains should contain no more than 3 to 4 molecules \cite{Russina2000,Gebremichael:PRE:2001}, most molecules are within 1.5 molecular diameters of an LC domain. In this regime, the entire system should relax on roughly the timescale for exchange of molecules into and out of LC domains, giving $\tau_{\alpha}\approx \tau_{\beta, JG}$. At $T<T_c$, $\Phi$ drops and LC domains become more scarce, so that it takes increasingly longer for $\beta_{JG}$ exchange events to accomplish $\alpha$ relaxation, leading to a bifurcation of these relaxation times. If the exchange events occurred randomly in space, the simple relation $\tau_\alpha=\tau_{\beta,JG}/\Phi$ would hold. The exponent, $\gamma$, relating changes in $\tau_\alpha$ and $\tau_{\beta,JG}$ arises because the LC domains are stringlike \cite{Gebremichael:PRE:2001}, so TC-LC exchange events will have nontrivial spatial correlations. It seems that essentially all fragility-related information is contained in $\gamma$, since $\left\langle \tau_{\beta,JG}\right\rangle$ appears to be nearly universal. 

In addition to TC-LC exchange, $\alpha$ relaxation should occur via a parallel intrinsic TC relaxation processes. The fact that we have accounted only for the latter but still obtain excellent fits indicates that the former is relatively slow, and this is consistent with the MCT result that the intrinsic $\alpha$ relaxation diverges \cite{Gotze:ZeitschriftFurPhysikBCondensedMatter:1987}. We suggests that intrinsic TC relaxation may not contribute significantly until higher temperatures, and that the anomalies observed near $T_A$ may be due changes in relative importance of TC and TC-LC exchange to $\alpha$ relaxation.

The mechanism we propose for $\beta_{JG}$ relaxation provides rationale for many known features of this relaxation process. For example, it justifies the close connection between $\beta_{JG}$ relaxation and translational diffusion \cite{richert2007enhanced,Yu:PhysicalReviewLetters:2012}, provides a  temperature-dependent activation energy as required by Dyre et al. \cite{Dyre:PhysicalReviewLetters:2003}, and provides a mechanism for the relevant PEL to be similar to that of the $\alpha$ relaxation \cite{Goldstein:JCPComm:2010}. It also provides for a connection between $\beta_{JG}$ and low-T heat capacity anomalies \cite{Johari:PhysicalReviewB:1986}, since the LC domains persist at low temperature \cite{Zhang:PhysicalReviewLetters:2011}. Further, the drop in $\Phi$ at low temperature explains the negative temperature dependence in the strength of $\beta_{JG}$ \cite{Kudlik:JMS:1999}.

Finally, we note that the presence of regions of extended mobility as evidenced here could give rise to an excess density of states, and thus could be related to the boson peak. On the other hand, the relationship would not be trivial, since the amplitude of the boson peak and the magnitude of $\Phi$ observed by us have very different temperature dependencies.

Based on our QENS data and literature cited herein, we have presented a simple two-state dynamic model that ties $\alpha$, $\beta_{JG}$, \& $\beta_{fast}$ relaxations, and translational diffusion to ps timescale dynamic heterogeneity in liquids. The model allows simple and quantitative calculation of timescales for each of these processes, and provides a molecular mechanisms for hopping and for the Johari-Goldstein $\beta$ process. We advocate a structural origin for the LC state related to packing frustration, and thus expect the proposed model to be generic and applicable to essentially all liquids.

\begin{acknowledgments}
We thank Jack Douglas, David Simmons, Mark Ediger, and Walter Kob for comments. We acknowledge funding from NIH/NIBIB under grant R01 EB006398-01A1. This work utilized facilities supported in part by the National Science Foundation under Agreement No. DMR-0454672. Official contributions of the National Institute of Standards and Technology. Not subject to copyright in the United States.
\end{acknowledgments}

\bibliography{TwoStateV3}

\end{document}